\documentclass[pdftex,onecolumn,epjc3]{svjour3}
\RequirePackage[T1]{fontenc}
\RequirePackage{fix-cm}
\RequirePackage{graphicx}
\RequirePackage{mathptmx} 
\RequirePackage{latexsym}
\RequirePackage{flushend}
\RequirePackage[numbers,sort&compress]{natbib}
\RequirePackage[colorlinks,citecolor=blue,urlcolor=blue,linkcolor=blue]{hyperref}
\RequirePackage{amsmath}

\newcommand{\ben}{\begin{eqnarray}}
\newcommand{\een}{\end{eqnarray}}
\newcommand{\be}{\begin{equation}}
\newcommand{\ee}{\end{equation}}
\newcommand{\n}{\label}
\newcommand{\no}{\noindent}

\newcommand{\ro}{\rho}
\newcommand{\om}{\omega}

\journalname{Eur. Phys. J. C}

\begin{document}
\title{ An exotic k-essence interpretation of interactive cosmological models}

\author{M\'onica Forte\thanksref{e1,addr1}}
\thankstext{e1}{e-mail:forte.monica@gmail.com}
\institute{Departamento de F\'isica, Facultad de ciencias Exactas y Naturales, Universidad de Buenos Aires, 1428 Buenos Aires, Argentina \label{addr1}}
\date{Received: date / Accepted: date}
\maketitle
\begin{abstract}
We define a generalization of scalar fields with non-canonical kinetic term which we call exotic k-essence or briefly, exotik.  These fields are generated by the global description of cosmological models with two interactive fluids in the dark sector and under certain conditions, they correspond to usual k-essences. The formalism is applied to the cases of constant potential and of inverse square  potential and also we develop the purely exotik version for the modified holographic Ricci type of dark energy (MHR), where the equations of state are not constant. With the kinetic function $F=1+mx$  and the inverse square  potential we recover, through the interaction term, the identification between k-essences and quintessences of exponential potential, already known for Friedmann-Robertson-Walker and Bianchi type I geometries. Worked examples are shown that include the self-interacting MHR and also models with crossing of the phantom divide line (PDL).
\end{abstract}


\section{Introduction}
\label{intro}
The cosmological  acceleration suggested by astrophysical data \cite{Perlmutter:1998np} can be explained by very different models,  among which, the simplest one is the $\Lambda$CDM \cite{Padmanabhan:2002ji}. A significant number of them reject the possibility of interaction between the  modeler components of dark matter (DM) and dark energy (DE), and perhaps for that reason, they fail to justify the coincidence problem. On the other hand, there is no fundamental reason to assume an underlying symmetry which would suppress the coupling. Whereas interactions between DE and normal matter particles are heavily constrained by observations (e.g. in the solar system and gravitational experiments on Earth), this is not the case for DM particles. Moreover, the possibility of DE-DM interaction is a phenomenon consistent not  only with recent Planck cosmological data \cite{Yang:2013ria} but  also it looks like theoretically possible when coupled scalar fields are considered \cite{Amendola:2003wa}, with many different types of interactions \cite{Cai:2004dk}. 
With respect to DE, among the several candidates to play its role, such as vacuum polarization, vector models, tachyons, Chaplygin gas, k-essences, Cardassian expansion, quasi-steady state cosmology and scalar-tensor models 
\cite{Parker:1999td,ArmendarizPicon:2004pm,Sen:2002nu,Bento:2003we,Chiba:1999ka,Freese:2002sq,Narlikar:2002bk,Amendola:2000uh,Bamba:2012cp}, the k-essence cosmology has received a lot of effort \cite{Kim:2004wj,Chimento:2003zf,Malquarti:2003nn} also playing both, the dark matter and dark energy roles \cite{Chimento:2004jm}. 
The k-essence would explain the coincidence problem because the transition between the tracker behavior during the radiation-matter domination and a cosmological constant-like behavior, appears to arise for purely dynamical reasons without fine-tuning necessity. 

In this work we try to connect both issues, cosmological systems of two interactive fluids on one side and unified models handled by a single field on the other side, with the intention to use the results of both fields of study, regarding them as different approaches to the same model.
The idea of describing a universe filled with two interactive fluids with constant equations of state $\omega_1$ and $\omega_2$, through a unified model driven by a single scalar field was first implemented in \cite{Chimento:2007da} and that produced a new type of field, dubbed exotic quintessence. This field, affected by an exponential potential, differed from the usual quintessence  because of the inclusion of the parameters $\omega_1$ and $\omega_2$ in the expressions of its evolution equation, its density of energy and its pressure.  For the values $\omega_1=1$ and $\omega_2=-1$, that is, when fluids are assumed to be stiff matter and vacuum energy, the usual quintessence was recovered. A special case of this field, when cold dark matter $\omega_1=0$ and vacuum energy $\omega_2=-1$ are considered, was used by Liddle et al. in \cite{Gao:2009me}. 
In this paper, we implemented the same idea through a new scalar field $\phi$ with non-canonical kinetic term, coming from a Lagrangian of the type  $L = -V(\phi)F(x)$ with $x=-(\dot \phi)^2$ \cite{Chimento:2003zf}.  We nicknamed exotic k-essence or exotik to this new class of non-canonical scalar, because of the new equations of motion that they fulfill, are reduced to a common k-essence under certain conditions.
 The connection with the work in \cite{Chimento:2007da} holds in the special case in which the quintessence approach is reduced to a common quintessence with exponential potential and the k-essence approach is reduced to a common  k-essence defined by the kinetic function $F(x)=1+mx$ and is affected by a potential of inverse square. In that case both models are described by the same scale factor and so they are geometrically equivalent. \cite{Chimento:2003ta},\cite{Chimento:2005ua}. The mechanism of generation of mathematical expressions for the exotic field, favors the role of one of the two interacting fluids. However, either of the two fluids is also able to perform the mathematical description. The defining choice will depend on the particular interest of each model because equations will include terms such as $(1+\omega_1)^{-1}$ if the key field is the second or $(1+\omega_2)^{-1}$ if the key field is the first.  In this paper we have in mind cosmological systems in which the fluid 1 is never considered as vacuum energy, and therefore, the logical choice is to take the fluid 2 as a key field generating factors $ (1+ \omega_1) ^ {- 1} $ that do not falsely  divergent the physical magnitudes into consideration. The next step is giving to  the key fluid (from now on fluid 2, playing the role of dark energy) a functional form that describes the specific description is intended to work. At this stage is where any anzats supported on specific physical models or described as combinations of physical quantities, appears. For example, the anzats taking linear combinations of density of energy  and pressure is formally equivalent to the expression used for the dark energy fluid in modified holographic Ricci (MHR) type models (if $\rho^{(MHR)}_2=2(\dot H+3AH^2/2)/(A-B)$  then  $\rho^{(MHR)}_2 =(1+A)/(A-B)\rho+1/(A-B)p$) \cite{Granda:2008tm}. The last step corresponds to express the above process  in terms of known elements, that is, expressions of the energy density and the pressure of a common k essence, $\rho=V(F-2xF_x)$  and  $p= -VF$.
The various possible exotik descriptions (each described by a particular ansatz),  are obtained by solving the equation of evolution for the key fluid affected by an interaction $Q$. Conversely, we can discover new types of interactions associated with known k-essences solving the same equation of evolution written in terms of known F and V. 

Our paper is organized as follows. In Section II we consider the general interacting two-fluid cosmological model and introduce the definition of exotic k-essence. In Section III  we gain deeper insight into the subject analyzing the representation with constant potential and with square inverse potential and also giving first integrals in both cases. In Sections IV we establish the equation which connects the exotik with an arbitrary interaction.  In sections V and VI we show worked examples for constant potential (including the modified holographic Ricci DE \cite{Nojiri:2005pu}), and for the square inverse potential (including the ratification of the equivalence between the linear k-essence $F= 1+mx$ and the quintessence with exponential potential) respectively.


\section{Exotik }


We consider a model consisting of two perfect fluids with an energy-momentum tensor $T_{ik} = T ^{(1)}_{ik}+ T ^{(2)}_{ik}$ . 
Here $T^{ (n)}_{ik}= (\rho_n + p_n)u_iu_k +p_n g_{ik}$ , where $\rho_n$ and $p_n$ are the energy density and the equilibrium pressure of fluid $n$ and $u_i$ is the four-velocity. 
Assuming that the two fluids interact between them in a spatially flat homogeneous and isotropic Friedmann-Robertson-Walker (FRW) cosmological model, the Einstein's equations reduce to:
\be
\n{01}
\small{3H^2= \ro_1 + \ro_2,}
\ee
\vskip -0.2cm
\be
\n{02}
\small{\dot\ro_1 + \dot\ro_2 +3H[(1+\om_1)\ro_1 + (1+\om_2)\ro_2]=0,}
\ee
\vskip 0.3cm
\no where $H = \dot a/a$ and $a$  stand for the Hubble expansion rate and the scale factor respectively and where we consider constant equations of state  $\om_i=(p_i/\ro_i)$ for $ i=1, 2$. 
 The whole equation of conservation (\ref{02}) shows the interaction between both fluid components allowing the mutual exchange of energy and momentum. Then, we assume an overall perfect fluid description with an effective equation of state, $\om = p/\ro = -2\dot H /3H^2 - 1$, where 
$p = p_1 + p_2$ and $\ro = \ro_1 +\ro_2$.  The dot means derivative with respect to the cosmological time and from Eqs. (\ref{01}), (\ref{02}) we get
\be
\n{03}
-2\dot H  = (1 + \om_1)\ro_1 + (1+ \om_2)\ro_2 = (1 + \om)\ro.
\ee
\vskip 0.3cm
As it  was done with an exotic canonical scalar field in  \cite{Chimento:2007da}, we propose  that the interactive 
system as a whole be represented by an exotic field $\phi$ (labeled by the potential function $V(\phi)$ and the kinetic function $F(x)$, $ x=-{\dot\phi}^2$, through the relationship
\be
\n{04}
(1 + \om_1)\ro_1 + (1+ \om_2)\ro_2 = -2V(\phi)xF_x(x), \  F_x= \frac {d F(x)}{d x}.
\ee
\vskip0.2cm
The global density of energy $\rho$ and  the global pressure $p=\om\ro$ are
\be
\n{05}
\small{\ro = \frac{1}{(1 + \om_1)}\left[-2V(\phi)xF_x(x)+\Delta\rho_2\right],}
\ee

\be
\n{06}
\small{p= -2V(\phi)xF_x(x)\frac{\om_1}{(1 + \om_1)}-\frac{\Delta}{(1 + \om_1)}\rho_2,}
\ee
\vskip0.1cm
\no where $\om_1\ne -1$, $\Delta=\om_1- \om_2$ and the field $ \phi $ satisfies  the conservation equation

\be
\n{07}
\small{\left[F_x+xF_{xx}\right]\ddot\phi+\frac{3}{2}(1 +\om_1)HF_x\dot\phi -\frac{V'}{2V}xF_x+\frac{\Delta}{4}\frac{\dot\rho_2}{V\dot\phi}=0 }
\ee
\no for \small{ $F_{xx}=d F_x/d x$}. 

The equations (\ref{05})-(\ref{07}) define the exotik field, a generalization of the k-essence field. \\
So many classes of exotiks exist as functional forms we adopted for $\rho_2$. The anzats
\be
\n{08}
\small{\rho_2(x)=V\left[\alpha F - 2x\beta F_x\right],} 
\ee
\no with constants $\alpha$ and $\beta$, can be thought as the general linear combination of the density of energy and  pressure of a common k-essence \footnote{$\rho_k = V(F-2xF_x)$ and $p_k = -VF$}

On the other hand, a similar expression arises naturally  in modified  holographic Ricci (MHR) dark energy models \cite{Granda:2008tm,Chimento:2011dw} if we think that the global model is driving by a k-essence \footnote{In this latter approach $\rho^{MHR}_2(x)=2[\dot H +3AH^2/2]/(A-B)$ and so the constants are related by  $A=\alpha/(\alpha-\beta)$ and $B=(\alpha-1)/(\alpha-\beta)$. Note that as  $\rho^{MHR}_2(x)=[-1-\om^{MHR} +A]\rho/(A-B)$, if $A>B$,  A represents the maximum possible value of the overall equation of state ($\om^{MHR}\leq A-1$) and therefore $A\leq 2$.}

This  proposal allows us to understand the reason of the name Exotik for this k-field as follow. The application of (\ref{08}) to the equations (\ref{05})-(\ref{07}) transforms them into
\begin{subequations}
\label{09}
\be
\label{9a}
\small{\ro = \frac{V}{(1 + \om_1)}\bigg(\alpha\Delta F-2x\left[1+\beta \Delta\right]F_x\bigg),}
\ee
\be
\n{9b}
\small{p= \frac{V}{(1 + \om_1)}\bigg (-\alpha\Delta F -2x\left[ \om_1-\beta \Delta\right]F_x \bigg),}
\ee
\no and
\be
\label{9c}
\small{\left[\left(2+(2\beta-\alpha)\Delta\right)F_x+2(1 + \beta\Delta)xF_{xx}\right]\ddot\phi }+\small{  3(1 + \om_1)HF_x\dot\phi  + }\small{ \frac{V'}{2V}\left[\alpha \Delta F- 2(1 +\beta\Delta)xF_x\right]=0.}
\ee
\end{subequations}

In this context, our exotik representation of the interacting system results in a habitual k-essence when $\alpha = (1+\om_1)/\Delta $ and $\beta=\om_1/\Delta$  for arbitrary pairs of fluids.  This motivates the name  exotik or exotic k-essence,  for the general case  where the  two independent parameters $\alpha$ and $\beta$ can not be considered superfluous or included into either the function F, or the potential V. 
Note that these very particular identifications for $ \alpha $ and $ \beta $ imply that $ \om_1 $ and $ \om_2 $ are constant and so it is not a recommended option when we are using a MHR fluid as DE.\footnote{{Instead of (\ref{05})-(\ref{07}) , the correct equations for the case of MHR  fluid as DE, are \\
$\rho^{MHR} = \frac{1}{A}\left[-2V(\phi)xF_x+(A - B)\rho_2\right],$\\
$p^{MHR}= -2VxF_x\frac{A - 1}{A}-\frac{(A - B)}{A}\rho_2,$\\
$\left[F_x+xF_{xx}\right]\ddot\phi+\frac{3}{2}AHF_x\dot\phi -\frac{V'}{2V}xF_x+\frac{(A - B)}{4}\frac{\dot{\rho_2}}{V \dot\phi}=0$\ \ \ \ \ \ }}   Because of in these holographic models always it is verified that $\om_2= (A-\om_1-1)\rho_1/\rho_2+B-1$ , a constant ratio of dark densities of energy arises, condition not necessarily true for general interactions.  In the general statement and from the equations (\ref{09}), the global equation of state $\om$ is read as
\be
\label{10}
\small{\om= - \frac{\alpha F - 2(\beta - \om_1/\Delta)xF_x}{\alpha F - 2(\beta + 1/\Delta)xF_x}.}
\ee


\section{Potentials and asymptotic behaviors}


The evolution equation (\ref{9c}) allows us to find the functional form of the exotik field once the potential and the kinetic function are given.  The choice of the potential is subject to the type of comparison in which we are interested. For example, the constant potential is needed to contrast with purely k-essences and the inverse square potential is used to collate with quintessences with exponential potentials.
\begin{itemize} 
\item{$V=V_0$} \\
In  the constant case, the equation (\ref{9c}) has the first  integral 
\be
\label{1aIntegrala}
\dot\phi ^{2+(2\beta-\alpha)\Delta } F_x^{1+\Delta\beta} =\frac{m_0}{a^{3(1+\om_1)}}
\ee
\no with $m_0$ being a constant of integration.
In turn, this expression leads to see a novel feature of the equation of state  (\ref{10}) of these models,
\be
\label{omegaV0}
\om=-\frac{(\beta\Delta-\om_1)+Ca^{\nu} FF_x^{\sigma -1}}{(\beta\Delta+1)+Ca^{\nu} FF_x^{\sigma -1}},
\ee
\no with the definitions $C=\alpha\Delta/2m_0^{2/(2-(\alpha - 2\beta )\Delta)}$, $\nu=6(1+\om_1)/(2-(\alpha - 2\beta )\Delta)$ and $\sigma=2(1+\beta\Delta)/(2-(\alpha - 2\beta )\Delta)$ for graphical economy.
It is known that the usual k-essences whose kinetic functions have a root \cite{Chimento:2004jm}, drive models with  a dust behavior at epoch around $t_0$ where $F(x(t_0))=0 $. These unified dark energy models are included here,  independently of the pair of fluids in interaction, when $ \beta = \om_1 /\Delta $. Actually, even if the kinetic functions have no  roots, the same behavior is observed if $a^{\nu} FF_x^{\sigma -1}\ll 1$ when $\nu > 0$. But our purely exotiks let go further and include the stages of radiative dominance at early times, properly choosing the representation with $\beta\Delta =(3\om_1 -1)/4$.   For $\nu > 0$ all these models show an accelerated behavior ($\om \rightarrow -1$) when  $a \rightarrow \infty $, providing appropriate behavior for our actual expanding universe.

\item{$V=V_0/\phi^2$} \\
For the inverse square case   $V=V_0/\phi^2$, equation (\ref{9c}) can be rewritten  in terms of the global equation of state $\om$ as
\be
\label{PotInverso}
\small{\ \ \ \ \ \ \dot\phi ^{u}\Bigg(\frac{1+\om}{\dot\phi ^{u}}\Bigg)^{.}+}\small{\Bigg(3H(1+\om)-\frac{2\dot\phi}{\phi}\Bigg)}\small{\Big(\lambda-1-\om \Big)=0,}
\ee
\no where $u=\alpha\Delta/(1+\beta\Delta)$ and $ \lambda=(1+\om_1)/(1+\beta\Delta)$.  This difficult equation has a first integral when we are dealing with common k-essence and more generally, asking $u = 1$, which is equivalent to having a single free parameter in (\ref{08}). This first integral can be written in three different ways
\begin{subequations}
\label{1aIntegralb}
\be
\label{1aIntegralb1}
\small{\frac{(1+\om)}{\dot\phi}\phi=\frac{2 }{3H}\Bigg( 1 + \frac{m_1}{a^{3\lambda}H} \Bigg),}
\ee
\be
\label{1aIntegralb2}
\small{\dot\phi F_x =\frac{\phi }{V_0}\Bigg( H + \frac{m_1}{a^{3\lambda}} \Bigg),}
\ee
\be
\label{1aIntegralb3}
\small{- V_0\dot H F_x =\Bigg( H + \frac{m_1}{a^{3\lambda}} \Bigg)^2,}
\ee
\end{subequations}
\no with $m_1$ a positive constant of integration. 

For the case $u=1$ is $\lambda=(1+\om_1)/(\alpha\Delta)$ and the equation (\ref{9a}) gives us the simple relation $VF=(\lambda-1-\om)\rho$ which allows to see that $\om=\lambda-1$ whenever F vanishes. Using (\ref{1aIntegralb2}), the global EoS (\ref{10})  is written as
\be
\label{omegacuadrInverso}
\small{\om=-\frac{ V_0V FF_x a^{6\lambda}-2(\lambda -1)\Bigg( m_1+a^{3\lambda}H\Bigg)^2}{ V_0VFF_x a^{6\lambda}+2\Bigg( m_1+a^{3\lambda}H\Bigg)^2}.}
\ee
From (\ref{omegacuadrInverso}) it follows that when F has a root, the global EoS is $\om= \lambda-1$ whatever the sign of $\lambda$ be. For times outside those cases, when $F(x)$ is monotone and non-zero, and considering that in cases of interest is $\lambda > 0$, the asymptotic behaviors are:  $\om= \lambda-1$ at early times ($a\rightarrow 0$) and 
$1+\om=2/(3F_x)$ at late times ($a \gg 1$).  The latter behavior generates a necessary condition to be satisfied by possible kinetic functions when the phantom regime is excluded. In that cases, from (\ref{10}) and $\lambda > 0$, the only admissible kinetic functions are the monotonous increasing F. 
\end{itemize}
The intention behind the attitude of not fixing $\alpha$ and $\beta$, is that the representation may still be used in the modified holographic case where these constants shape the upper limit of the value of the global barotropic index. Also, as we will show in the next section, some of them do not allow for adequate representation while others facilitate the resolution of the problem.


\section{The interactions}


The above results are quite general and apply to any kinetic function $F(x)$, but the particular choice of the function will be determined by the interaction $Q$ amending the evolution of both fluids.  We define the interaction $Q$, through the partition of the global conservation equation (\ref{02}), as

\begin{subequations}
\label{11}
\be
\n{11a}
\dot\rho_1 + 3H(1+\om_1)\rho_1 = -3HQ,
\ee
\be
\n{11b}
\dot\rho_2 + 3H(1+\om_2)\rho_2 = 3HQ.
\ee
\end{subequations}

Then, the expressions (\ref{08}), (\ref{9c}) and (\ref{11b}) let us write the equation that must be fulfilled by the kinetic function $F(x)$ once the interaction $Q(V,F)$ and the potential $V$ are fixed.
\be
\n{FconVgeneral}
\small{3H\Bigg[\bigg(Q/V-(1+\om_2)(\alpha F - 2xF_x\beta)\bigg)(2M-N\Delta)}\small{+2(1+\om_1)NxF_x \Bigg ]} = \small{2\alpha\frac{\dot V}{V}\Big( MF-xF_x^2 \Big),}
\ee
\no with $ M=F_x+xF_{xx} $ and $ N=(\alpha-2\beta)F_x-2xF_{xx}\beta $. 
The expression $Q(V,F)$ means that the interaction, often expressed as a function of $\rho$  and its derivatives, should be given using equations (\ref{01}), (\ref{02}), (\ref{08}) and (\ref{09}).\footnote{\textrm{The following  expressions are useful to express the interactions} \\
$\rho=\frac{V}{(1+\om_1)}\big(\alpha\Delta F-2x\left(1+\beta \Delta\right)F_x\big),$ \\
$\rho'=\frac{\dot\rho}{3H}=2xF_xV, $\\
$\rho_1 =-\frac{(1+\om_2)\rho+\rho'}{\Delta},$  \ \  $\rho_2 =\frac{(1+\om_1)\rho+\rho'}{\Delta},$ \\
$\rho''=\frac{2V}{3H}\big((xF_x)^{.}+xF_x\frac{\dot V}{V}\big). $}

The extremely complex equation (\ref{FconVgeneral}) is made simple in several important cases as those where $V=V_0$, constant.

\section{ Purely exotiks} 


The equation (\ref{FconVgeneral}) for $V=V_0$, is reduced to
\be
\n{18a}
\big(Q/V_0-(1+\om_2)(\alpha F - 2xF_x\beta)\big)(2M-N\Delta) + 2(1+\om_1)NxF_x = 0,
\ee
\vskip0.2cm
\no which is a highly nonlinear equation for F.  \\
However, the change of variables $\scriptsize{\zeta=\int {\rho_x/(2xF_xV_0)}dx}$ and $\scriptsize{\rho(x)=V_0\big(\alpha\Delta F-2x\left[1+\beta \Delta\right]F_x\big)/(1 + \om_1)}$  lets us to obtain the more simple  differential equation for $\rho$,  
\be
\n{18b}
\rho'' + (2+\om_1+\om_2)\rho'+ (1+\om_1)(1+\om_2)\rho=Q\Delta
\ee
\no with $\rho'=d\rho/d\zeta$ and $\rho''=d^2\rho/d\zeta^2$. This is the already known  source equation for the energy density described in \cite{Chimento:2009hj}.
\vskip-1.5cm
\begin{itemize}     
\item
\emph{Examples $Q \rightarrow F $}
 
\begin{itemize}     
\item
\emph{$\Lambda$CDM}\\
The trivial case $Q=0$, $\om_1=0$ and $\om_2=-1$, which corresponds to the $\Lambda$CDM model can be represented by the kinetic function $\small{F(x)=F_0 + F_1(-x)^{\frac{\alpha}{2\beta}}}$ with $F_0$ and $F_1$, two constant of integration. Using the first integral $\small{(-x) ^{1+\beta-\alpha/2} F_x^{1+\beta} =m_0/a^3}$ we recover the known expression $\rho=\rho_{01}/a^3 + \rho_{02}$ with $\rho_{01}=2 V_0 m_0(-2\beta)^{\beta}(F_1\alpha)^{-\beta}$ and $\rho_{02}=\alpha V_0 F_0$. This case shows that in general, a purely k-essence will not be a correct interpretation of that interaction because it leads to a constant density of energy.  Therefore, in principle, we must consider the two parameters $\alpha$ and $\beta$ not fixed.  
\vskip0.3cm
\item
\emph{CDM} and \emph{MHRDE}\\
This is a very interesting case because, although the dark energy EoS $\om_2$ is not a constant, the option for the density of dark energy $\rho^{MHR}_2=(2\dot H+3AH^2)/(A-B)=(A-\gamma)\rho/(A-B)$ relates $\om_1$ and $\om_2$ with $A$ and $B$ through the expression $\om_1\rho_1+\om_2\rho_2=A\rho_1+B\rho_2$. Taking $\om_1=0$ we define the modified interaction $Q_M$ through 
\be
\n{11bHolografico}
\ \ \ \ \ \ \ \ \ \ \ \ \ \dot\rho_2 + 3HB\rho_2 = 3HQ_M=3H(Q+(1-A)\rho_1)
\ee
\no and so, the equation (\ref{18a}) changes to 
\be
\n{18aHolografica}
\Big(Q_M-B\frac{(AF - 2xF_x(A-1))}{(A-B)}\Big)(2M-N(A-B))+ 2xF_xV_0 AN = 0.
\ee
The expression of $\rho^{MHR}_2$ shows that there is a perfect agreement with the anzats (\ref{08}) for $\alpha=A/(A-B)$ and $\beta=(A-1)/(A-B)$ and so the exotik function is a common k-essence. We apply this to the null interaction $Q=0$, which is equivalent to replacing $Q_M=Q+(1-A)\rho_1$ in (\ref{18aHolografica}). The solution is $F(x)=(F_0+F_1\sqrt{-x})^{B/(B-1)}$ and lets writing the densities of energy $\rho^{MHR}=b_1 a^{-3}+b_2 a^{-3B}$ and $\rho^{MHR}_2=((A-1)/(A-B))b_1 a^{-3}+b_2 a^{-3B}$. So, the purely exotik includes the modified holographic Ricci DE model with cold dark matter, where it can be seen that, even when Q is null, the dark energy component is far from remaining independent of the CDM. This is a consequence of the "holography" of the model introduced into the expression of $\rho^{MHR}_2$\cite{Chimento:2011dw}. In other words, the MHR fluid, is always a self-interacting component.
\vskip0.3cm

\item
\emph{ $Q=\tau \rho'/\Delta$}\\
Halfway between choosing none or both parameters, sometimes, set just one makes it easier to reach the goal. Select which and how to fix it, it is evident through the process of solving the equations.  For example, in the case of interaction proportional to $\rho'$,  $Q=\tau \rho'/\Delta$ it is highly advisable to set $\beta=-1/\Delta$. Then, we obtain the first integral  $\small{(-x) ^{-\alpha \Delta/2} =m_0a^{-3(1+\om_1)}}$ and the appropriate exotik 
\be
\n{raices1}
\ \ \ \ \ \ \ \ \ \ \ \ \ \ \ F(x)= F_1(-x)^{\frac{-\alpha\Delta}{2(1+\om_1)} n_+}+ F_2(-x)^{\frac{-\alpha\Delta}{2(1+\om_1)}n_-},
\ee
\no  with 
\be
\n{raices2}
2n_{\pm}=\om_1 + \om_2+2-\tau \pm\sqrt{(\om_1+\om_2+2-\tau )^2- 4(1+\om_1)(1+\om_2)}.
\ee
The corresponding  global density of energy $$\rho=\rho_{01}a^{-3 n_+}+ \rho_{02}a^{-3 n_-}$$ and global EoS
$$\ \ \ \ \ \ \ \ \ \ \ \om=-1+\frac{n_+  F_1+ n_- F_2a^{3 (n_+ - n_-)}}{F_1+ F_2a^{3 (n_+-n_-)}}$$ show that the global behavior is determined by the relationship between the strength of the coupling $\tau$ and the EoS of the considered fluids. As $(n_+ - n_-)$  is always positive, the asymptotic values are $\om \rightarrow n_+ -1$ at early times and $\om \rightarrow n_- -1$ at late times.  The demeanor of the model  is like a quintessence or phantom according to whether $\tau <\om_1 + \om_2+2$ or $\tau >\om_1 + \om_2+2$, but there is no crossing of the phantom divide line (PDL).  The option $\tau =\om_1 + \om_2+2$ is only admissible in the case of a cosmological constant ($\om_2=-1$) and there, the effect of this interaction is to freeze the overall density of energy. Could this interaction be a mechanism to freeze the densities of elementary particles in the primordial times, assuming the existence of a cosmological constant?

\end{itemize}   
It is important to note that $ \alpha $ and $ \beta $ determine the representation and we must work within it.  However, the results of the corresponding models do not depend on these parameters. This should be clear in the next section where we go from F to Q,  which is expressed through the magnitudes given by the particular representation. The effects are studied within the representation but not depend on it.

\item
\emph{Examples $F \rightarrow Q $}

\begin{itemize}     
\item
An interesting example in the representation of common k-essence corresponds to the Chimento function  $F_{Ch}(x)=\frac{1}{V_0(2n-1)}\left[2nn_0\sqrt{-x} - (-x)^{n}\right]$, $n\ne0$ and $n\ne1/2$. 
Using (\ref{18a}) we obtain the associated interaction $Q_{Ch}$
\be
\n{QChim}
Q_{Ch}\Delta=(1+\om_1)(1+\om_2)\rho+(\om_1+\om_2+1)\rho'+\frac{1}{2n}\frac{\rho'^2}{\rho}.
\ee
$F_{Ch}$ have a root at $x_r=-(2nn_0)^{2/(2n-1)}$ and an extreme at $x_{ex}=-n_0^{2/(2n-1)}$. The equation (\ref{omegaV0}), now with $\sigma=1$,  lets us know that the cosmological model driven by $Q_{Ch}$ has an era with like dust behavior around $a_{root}^3=m_0^{1/(1+\om_1)}/nn_0$  and a cosmological constant behavior at late times, regardless of the duet of fluid considered. The corresponding total density of energy is easily expressible from (\ref{9a}) y (\ref{1aIntegrala}) as
$$ \ \ \ \ \rho_{\small{Ch}}=V_0\big\{ n_0 + \frac{(2n-1)m_0^{1/(1+\om_1)}}{n a^3}\big\}^{\frac{2n}{2n-1}}.$$ 
Moreover, writing (\ref{omegaV0}) as
$$\ \ \ \ \ \ \ \ \ \ \ \ \ \ \om=- \frac{v(v-1)(1+(2n-1)v)^{\frac{1}{2n-1}}}{2n^2n_0^{\frac{2(n-1)}{2n-1}} + v(v-1) (1+(2n-1)v)^{\frac{1}{2n-1}}},$$
with $v=(a/a_{root})^3$ it can be seen that  these models present a dust like behavior at early times. Also, they look as a $\Lambda$CDM models with no interaction at all for $n>>1$ in which case, the dust-like behavior at the root of $F$ coincides with the dust-like behavior at the epoch $a << 1$.
\vskip0.5cm
    
\item
Another interesting example arises, again in the common k-essence representation,  $\alpha = (1+\om_1)/\Delta $ and $\beta=\om_1/\Delta$, when  we are really dealing with  a Chaplygin gas, through the kinetic function $F(x)=\sqrt{1+x}$.  From (\ref{18a}) we can express the interaction as
\be
\ \ \ \ \ \ \ \ \ \ \ \ \ \ \ \ \ \ \ Q\Delta=\rho''\Big( \frac{(1-\om_1)\rho+\rho'}{2\rho+\rho'}\Big)+(1+\om_2)\rho'+(1+\om_1)(1+\om_2)\rho.
\ee
and the corresponding density of energy $\rho$ is obtained directly from (\ref{9a}) and (\ref{1aIntegrala}), as
$$\rho=V_0 \sqrt{1+\frac{4m_0^{2/(1+\om_1)}}{a^6}}.$$
\end{itemize}     
\end{itemize} 

\section{Exotiks with inverse square potential} 

For cases with inverse square potential $V=V_0/\phi^2$ the expression (\ref{FconVgeneral}) can be written as
\be
\n{FconVinverso}
\small{\bigg(Q-(1+\om_2)\rho_2\bigg)(2M-N\Delta)+ 2(1+\om_1)NxF_xV}= \small{-2\alpha V\Big( MF-xF_x^2 \Big)\sqrt{\frac{2(1+\om)}{3F_xV_0}}.}
\ee

\begin{itemize}     
\item 
$\boldmath{F(x)=1+mx}$\\
As a first example we get the interaction term in the general representation (with  $\alpha$ and $\beta$ free) for the simplest kinetic function that supports a non-constant equation of state, $F(x)=1+mx$, with $m$ constant. In this case (\ref{FconVinverso}) is written as
\be
\n{QVinverso}
\ \ \ \ \ \ \ \ \ Q=(1+\om_2)\rho_2+\rho_2'.
\ee 
The usual k-essence version ($\alpha=(1+\om_1)/\Delta$, $\beta=\om_1/\Delta$) of (\ref{QVinverso}) allows us to show, through the interaction term, that the models driven by k-essence with inverse square potential and kinetic function $F(x)=1+mx$, and those dominated by a quintessence $\varphi$ with exponential potential $U(\varphi)=U_0 exp((-\sqrt{2/mV_0}\varphi)$ share the same geometry. This result was obtained in FRW \cite{Chimento:2003ta}  and also in Bianchi I cosmologies \cite{Chimento:2005ua}.
To test this feature, we must consider that the exotic quintessence used in \cite{Chimento:2007da} becomes a usual one when $\om_1=1$ and $\om_2=-1$ and so, it is the scalar representation of two fluids with the interaction $Q=-\rho_2\sqrt{2(1+\om)/(3mV_0)}$. Replacing the corresponding values of the equations of state at the usual k-essence version of (\ref{QVinverso}) we find the same equivalence because the interaction terms in both models agree. 
Moreover, note that the (\ref{QVinverso}) vanishes identically the source equation \cite{Chimento:2009hj} so that it is not possible to solve the system using that method. But here, using $F=1+mx$ in (\ref{1aIntegralb3}) we obtain the equation $mV_0\dot H+ H^2+2m_1 H a^{-3}+m_1^2 a^{-6}=0$ whose solutions for the factor of scale and the k-essence field have already been given in \cite{Chimento:2003ta}.

\item
$\boldmath{F(x)=\sqrt{-x}+x}$\\
The simplified Chimento function $F(x)=\sqrt{-x}+x$, formerly used for constant potential and now for inverse square potential, shows the power of this interpretation in usual k-essence representation  to resolve the evolution of  systems with two arbitrary fluids controlled by a nonlinear interaction. From (\ref{FconVinverso}) this function leads to the highly nonlinear interaction
\be
\n{QChimn1}
Q_{\nu}\Delta=(\om_2-1)(\om_1+1)\rho + (\om_2-1)\rho'+ (2\rho+\rho')(\om_1-\nu)+\frac{(2\rho+\rho')^2}{\rho},
\ee
\no with $\nu=(3V_0)^{-1/2}$ a positive constant. The general solution for the total density of energy in interactive systems with this interaction is 
$\rho=(\rho_{01}a^{-3}+\rho_{02}a^{-3\nu})^2$. Since this kinetic function transforms the equation (\ref{9a}) in $3H^2=\rho=-xV=V_0\dot\phi^2/\phi^2$, the exotik field is $\phi=\phi_0a^{3\nu}$, $\phi_0=\phi(a=1)$. The corresponding global EoS, $\om=-1-\rho'/\rho$ can be written as
\be
\n{EosChimN1}
\om=1-\frac{1}{\nu\phi_0(\rho_{01}a^{3(\nu-1)}+\rho_{02})},
\ee
\no and shows that the global system interpolates between a stiff fluid and a fluid with $\om_{asym}=1-1/\nu \phi_0 \rho_{02}$ in either direction, depending on whether $\nu $ is less than or greater than 1. The interesting case is $\nu <1 $, with stiff behavior at early times, that even admits the crossing of the phantom divide line, at late times.

\end{itemize}


\section{Conclusions}

We have described cosmological systems composed by two interactive fluids in the dark sector, with constant equations of state $\om_1$ and  $\om_2$, by means of unified models that are controlled by non-canonical versions of a scalar field. The description produces a generalization of k-essence field we have called  exotic k-essence or exotik.
There are many kinds of exotiks according to the expression proposed for the fluid representing the dark energy $\rho_2$ . The anzats used here is a linear combination of the density of  energy and the pressure of usual k-essences, parameterized by the constant values $\alpha$ and $\beta$. Then, the different representations or exotiks are labeled by the particular choice of the potential $ V $ and by those parameters, noting that the exotik description is reduced to a common k-essence when $\alpha=(1+\om_1)/(\om_1-\om_2)$ and $\beta=\om_1/(\om_1-\om_2)$. 
The derivation of the equations for the exotik field requires that the equations of state are constant, which is not the case in the modified holographic Ricci model. However, there is a simple relationship between the two EoS and the constants $A$ and $B$ in the DE expression $\ro_2^{MHR}$  and so the overall description can be carried out according to the process indicated at the footnotes in the end of bibliography. 
The choice constant $V =V_0$ simplifies the evolution equation for exotik field and its first integral is useful to observe novel features compared to the usual k-essence. In the latter case it is proven that the kinetic functions that have a root $F(x(t_0))=0$ have a dust behavior in the epoch around $t_0$ and also the same happens when $a^{\nu} FF_x^{\sigma -1}<<1$, that is, at early times, for $\alpha \Delta >0$. The exotik includes these cases but also allows us to include the time of radiation domination. Choosing the appropriate representation $\beta=(3\om_1-1)/4(\om_1-\om_2)$ and $2\alpha (\om_1-\om_2)<3(1+\om_1)$ the model interpolates between the radiation like era at early times and $\om \rightarrow -1$ at late times.
The choice $V =V_0 \phi^{-2}$ is also an interesting representation because, even discarding solutions with stationary global EoS, there are first integrals for the evolution equation of common k-essence and more generally for representations with $\alpha\Delta=1+\beta\Delta$, that is only one free parameter.
The ability to model the cosmological behavior at early times still exists in this more complicated pose although here the realistic models should be conducted with strictly increasing kinetic functions if we want to avoid the phantom regime.
Having chosen the representation, $\alpha$ and $\beta$, the particular kinetic function $F$ or the interaction $Q$ are defined as the solution of equation (\ref{FconVgeneral}) for a given potential. 
For $V =V_0$ we have shown the "self-interaction" of the modified holographic Ricci dark energy models and have found the exotik kinetic function in the case $Q=\tau \rho'/\Delta$ where we have found that the phantom or non-phantom regime depends on the strength of the coupling  constant $\tau$ not allowing the crossing of the phantom divide line. Also we have found the interaction associated with the Chimento function and the associated with a purely Chaplying gas. Interestingly, the two complex interactions have an easy resolution from the point of view of exotik representation.  For $V =V_0 \phi^{-2}$, we have found that the observed equivalence (in FRW and in Bianchi I backgrounds) between usual k-essence with $F(x)=1+mx$ and usual quintessences  with exponential potential is still observed at the level of associated interactions. Finally, we found the  interaction $Q_{\nu}$ corresponding to a simplification of the Chimento function, for which we express the exotik field and the global EoS as functions of the factor of scale showing that for $\nu<1$ the models admit the crossing of the PDL, at late times. 
These exotik representations allow leverage all the work done with  k-essence unified models giving them an interactive justification inside the dark sector.

\begin{acknowledgements}
The author acknowledges the comments and suggestions of the anonymous referee who allows the clarification of the  work.
\end{acknowledgements}


\end{document}